\documentclass[aps,pra,preprint,groupedaddress,twocolumn,10pt]{revtex4-1}
\usepackage{graphicx}
\usepackage{epsfig}
\usepackage{epstopdf}
\usepackage{subfigure}
\usepackage{appendix}
\usepackage{amsmath,xfrac,amssymb}
\begin{document}
\title{Optical switching and bistability in four-level atomic systems}
\author{Pardeep Kumar}

\email[]{pradeep.kumar@iitrpr.ac.in}
\author{Shubhrangshu Dasgupta}

\affiliation{Department of Physics, Indian Institute of Technology Ropar, Rupnagar, Punjab 140001, India}

\date{\today}

\begin{abstract}
We explore the coherent control of nonlinear absorption of intense laser fields in four-level atomic systems. For instance, in a four-level ladder system, a coupling field creates electromagnetically induced transparency (EIT) with Aulter-Townes doublet for the probe field while the control field is absent. A large absorption peak appears at resonance as the control field is switched on. We show how such a large absorption leads to optical switching. Further, this large absorption gets diminished and a transparency window appears due to the saturation effects as the strength of the probe field is increased.  We further demonstrate that the threshold of the optical bistability can be modified by suitable choices of the coupling and the control fields. In a four-level Y-type configuration, the effect of the control field on saturable absorption (SA) and reverse saturable absorption (RSA) is highlighted in the context of nonlinear absorption of the probe field. We achieve RSA and SA in a simple atomic system just by applying a control field.
\end{abstract}

\pacs{}

\maketitle

\section{Introduction}
Since the invention of laser, the quantum coherence  has played a vital role in controlling the nonlinear optical properties of an atomic medium \cite{scully1997, ficek2007}. The linear optical effect is attributed to the linear relationship between the input and output intensity. When the incident irradiance becomes large enough, nonlinear optical effect comes into play \cite{boyd2008}. Quantum coherence 
helps in enhancing the efficiencies of nonlinear optical processes (characterized by a complex third order susceptibility), thereby eliminating the linear absorption even at low light powers \cite{field1990,hau1998,xiao2001,kang2003,sheng2011}. While the real part of the third order susceptibility (the Kerr nonlinearity) can produce significant cross-phase modulation \cite{schmidt1996,schmidt1998}, the imaginary part may lead to photon switching \cite{harris1998, yan2001}.
Further, quantum coherence can produce coherent population oscillations and sub-linewidth transmission resonances in two-level \cite{boyd1981} as well as in multilevel systems \cite{kumar2013}.  This happens as an effect of temporal modulation of the population difference between the ground and excited state.
Besides transparency, electromagnetically induced absorption (EIA) may occur due to transfer of two-photon coherence between the degenerate excited levels to the degenerate ground levels \cite{akulshin1997, lezama1999, taichenachev1999}.  For the observation of the EIA, the multi-level systems, especially the four-level N-type systems \cite{kang2004, salloum2010} are the promising candidates, as they require a control field with low power. Further, two-photon coherence in ladder-type energy-level configuration \cite{mohapatra2007,kumar2009,liu2012} exhibits electromagnetically induced transparency (EIT) and two-photon absorption (TPA) for a weak probe laser, in presence of a strong coupling laser. This TPA phenomena in the ladder systems can be dramatically modified by quantum coherence and it can make an absorptive medium transparent to the probe field \cite{kaiser1961}. Such possibility has been originally proposed by Agarwal and coworkers \cite{agarwal1996} and demonstrated in \cite{gao2000}. Besides TPA, a large absorption at resonance can also arise due to three-photon coherence, termed as three photon electromagnetically induced absorption \cite{moon2014,moon2015}.

All the above nonlinear phenomena assume a probe field, strength of which is such that it is enough to consider the coherence only up to a finite order of the probe field (i.e., only a finite order of susceptibility, say, up to third order). However, when the intensity of the probe field is increased, one needs to investigate the response of the medium for all orders of the probe field that gives rise to several interesting optical effects, namely optical bistability (OB). This refers to possibility of two stable output fields for the same input field in an optical feedback network. The  nonlinear interaction between a collection of atoms and the field mode along with the feedback inside an optical cavity leads to such a bistable behavior.
Based on the response of the optical feedback, the bistable device can be used as an optoelectronic component, viz., an optical differential amplifier \cite{abraham1982,gibbs1985,lugiato1984}.  The optical bistability has been extensively studied in two-level \cite{maccall1974,gibbs1976,grant1982} as well as in multi-level systems \cite{harshawardhan1996,joshi2003,babu2013} due to its application in ultrafast all-optical switches \cite{gauthier2005}. Harshawardhan and Agarwal \cite{harshawardhan1996} have proposed that by using quantum interference induced by control fields, one can decrease the threshold of bistability. This suggests that nonlinear effect becomes dominant even at a low input intensity.

We further investigate the role of coherence into other nonlinear optical phenomena,  namely saturable absorption (SA) and reverse saturable absorption (RSA) \cite{sutherland2003}. The SA corresponds to the decrease of the ground state absorption of light as the input intensity increases. On the contrary, the RSA is associated with the excited state absorption and as a consequence, the absorption inside the medium increases with the intensity of the input field.
The RSA has been demonstrated in various compounds \cite{tutt1992,chen1999,vivien1999,zhao2010,lim2011} and has found application in \textit{optical limiting}, viz., protection of electro-optical sensors or human eyes from intense laser pulses \cite{zhou2010}.

In this paper, we explore the effect of a control field on the nonlinear absorption characteristics of a probe field in four-level atomic systems.  We consider all orders of the probe field in our analysis. The effect of atomic coherence as a function of all the fields is investigated. We show that while in the absence of the control field, the system exhibits transparency for the probe field,  the nonlinear excitation is enhanced as the control field is introduced in the uppermost transition of a four-level ladder system. This enhancement of the nonlinear absorption leads to the absorptive optical switching \cite{harris1998,yan2001} in which the probe absorption can be switched on and off by a control field. We present numerical results to describe the optical switching in terms of a Gaussian intensity profile of the probe field. We also describe the absorption characteristics for the atomic medium in a unidirectional ring cavity, leading to the optical bistability. In our model, the threshold of the OB can be controlled both by the control and the coupling field. Further, via coupling the first excited state with a metastable state by a control field (so as to form a Y-type configuration),  we show how the SA and RSA can be coherently manipulated, when both the coupling field and the probe field are of the same polarization and the same intensity. Note that here we demonstrate the RSA effect in a simple \textit{atomic system} thanks to the coherent \textit{control field}.

The structure of the paper is as follows. In Sec. II, we discuss the four-level ladder model configuration along with its relevant equations. In Sec. III, we describe how nonlinear absorption and optical bistability can be coherently controlled in such a system. Further, in Sec. IV, the SA and RSA effects are highlighted in a four-level Y-type system. We conclude the paper in Sec. V.

\section{Model}
\begin{figure}[!ht]
\includegraphics[scale=0.23]{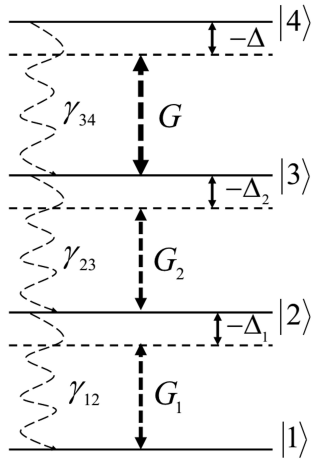}
\caption{Schematic of energy-level structure of a four-level ladder system. The coupling field of Rabi frequency $2G_{1}$ and the control field of Rabi frequency $2G$ induce the transitions $|1\rangle\leftrightarrow|2\rangle$ and $|3\rangle\leftrightarrow|4\rangle$, respectively, while a probe field of Rabi frequency $2G_{2}$ interacts with $|2\rangle\leftrightarrow|3\rangle$.}
\label{fig1}
\end{figure}
To explore the nonlinear absorption characteristics of the probe field and optical bistability, we consider a four-level ladder scheme \cite{sandhya2007}, as shown in Fig. \ref{fig1}, that comprises a ground state $|1\rangle$ and three excited states $|2\rangle$, $|3\rangle$ and $|4\rangle$, in increasing order of frequency.  The coupling field $\vec{E}_{1}=\hat{\pi}\varepsilon_{1}e^{-i\omega_{G_1}t}+c.c.$  $\left( \mbox{the probe field}~\vec{E}_{2}=\hat{\sigma}_{+}\varepsilon_{2}e^{-i\omega_{G_2}t}+c.c.\right)$ with a Rabi frequency $2G_{1}=\frac{2\vec{d}_{21}\cdot \hat{z}\varepsilon_{1}}{\hbar}$ $\left(2G_{2}=\frac{2\vec{d}_{32}\cdot\hat{\sigma}_{+}\varepsilon_{2}}{\hbar}\right)$ drives the $|1\rangle\leftrightarrow |2\rangle$ $\left(|2\rangle\leftrightarrow |3\rangle\right)$ transition, where $\varepsilon_{i}~\left(i=1,2\right)$ represents the amplitude of the field and $\vec{d}_{ij}$ is the electric dipole moment matrix element between the levels $|i\rangle$ and $|j\rangle$. Here, $\omega_{G_{1}}$ ($\omega_{G_{2}}$) denotes the frequency of the coupling (probe) field. We apply a control field $\vec{E}_{c}=\hat{\sigma}_{-}\varepsilon_{c}e^{-i\omega_{c}t}+c.c.$  of frequency $\omega_{c}$ and amplitude $\varepsilon_{c}$ on the transition $|3\rangle\leftrightarrow|4\rangle$. Let $2G=2\left(\frac{\vec{d}_{43}\cdot\hat{\sigma}_{-}\varepsilon_{c}}{\hbar}\right)$ be the Rabi frequency of the control field $\vec{E}_{c}$. The transitions $|1\rangle\leftrightarrow|3\rangle$, $|2\rangle\leftrightarrow|4\rangle$, and $|1\rangle\leftrightarrow|4\rangle$ are electric dipole forbidden.

The Hamiltonian for the system under consideration  in the dipole approximation  can be written as
\begin{equation}
\begin{array}{lll}
\hat{H}&=& \hbar \left[\omega_{21}|2\rangle\langle 2|+\omega_{31}|3\rangle\langle 3|+\omega_{41}|4\rangle\langle 4|\right]\\ && -\left[(\vec{d_{21}}\vert2\rangle\langle1\vert+H.c.).\vec{E_1}\right] -\left[(\vec{d_{32}}\vert3\rangle\langle2\vert+H.c.).\vec{E_2}\right]\\ &&-\left[(\vec{d_{43}}\vert4\rangle\langle3\vert+H.c.).\vec{E_c}\right]\;.
\end{array}
\label{eq1}
\end{equation}

Here $\hbar\omega_{\alpha\beta}$ is the energy difference between the levels $|\alpha\rangle$ and $|\beta\rangle$ and zero of energy is defined at the level $|1\rangle$. The dynamical evolution of the system can be described by Markovian master equation and the relevant density matrix equations obtained for the four level ladder system are given in the Appendix [see Eq. (\ref{eqa1})]. Note that the Eqs. (\ref{eqa1}) consider the mutual effect of atomic coherence and all the relevant fields.

The induced polarization at a frequency $\omega_{G_{2}}$ on the transition $|2\rangle\leftrightarrow|3\rangle$ will be obtained from the off-diagonal matrix element $\tilde{\rho}_{32}$:
\begin{equation}
  P\left(\omega_{G_{2}}\right)=\mathcal{N}\vec{d}_{23}\tilde{\rho}_{32}\;,
\label{eq2}
\end{equation}
where, $\mathcal{N}$ is the number density of the medium. Below, we study the propagation of the field through the medium using Maxwell's equations. Using the slowly varying envelope approximation, the equations for field propagation can be expressed as

 \begin{align}\label{eq3}
\left(\frac{\partial}{\partial z}+\frac{1}{c}\frac{\partial}{\partial t}\right)G_{1}&=i\eta_{12}\tilde{\rho}_{21}, ~~~ \eta_{12}=3\lambda_{12}^{2}\mathcal{N}\gamma_{12}/4\pi\;,\nonumber\\
\left(\frac{\partial}{\partial z}+\frac{1}{c}\frac{\partial}{\partial t}\right)G_{2}&=i\eta_{23}\tilde{\rho}_{32}, ~~~\eta_{23}=3\lambda_{23}^{2}\mathcal{N}\gamma_{23}/4\pi\;,\\
\left(\frac{\partial}{\partial z}+\frac{1}{c}\frac{\partial}{\partial t}\right)G &=i\eta_{34}\tilde{\rho}_{43},~~~\eta_{34}=3\lambda_{34}^{2}\mathcal{N}\gamma_{34}/4\pi\;\nonumber.
\end{align}
where, $\eta_{ij}$ and $\lambda_{ij}$ are the coupling constant and wavelength for $|i\rangle\leftrightarrow|j\rangle$ transition, respectively.
\section{Results}
For the numerical studies, we consider relevant transitions in $^{23}$Na atoms \cite{fischer2006}, where $|1\rangle=$3S$_{1/2}$, $|2\rangle=$3P$_{1/2}$, $|3\rangle=$3D$_{3/2}$, and $|4\rangle=$8P$_{1/2}$. The respective transition wavelengths and decay rates are $\lambda_{12}$=594 nm and $\gamma_{12}\approx$ 9 MHz (for $|1\rangle\leftrightarrow|2\rangle$ transition), $\lambda_{23}$=821 nm and $\gamma_{23}\approx$ 6 MHz (for $|2\rangle\leftrightarrow|3\rangle$ transition), $\lambda_{34}$=986 nm and $\gamma_{34}\approx$ 0.005 MHz (for $|3\rangle\leftrightarrow|4\rangle$ transition). Further, for numerical results, we set $\gamma_{12}\approx\gamma_{23}\approx\gamma$, unless stated otherwise.
\subsection{Control of transmission}
\begin{figure}[!ht]
\begin{center}
\begin{tabular}{cc}
\includegraphics[scale=0.4]{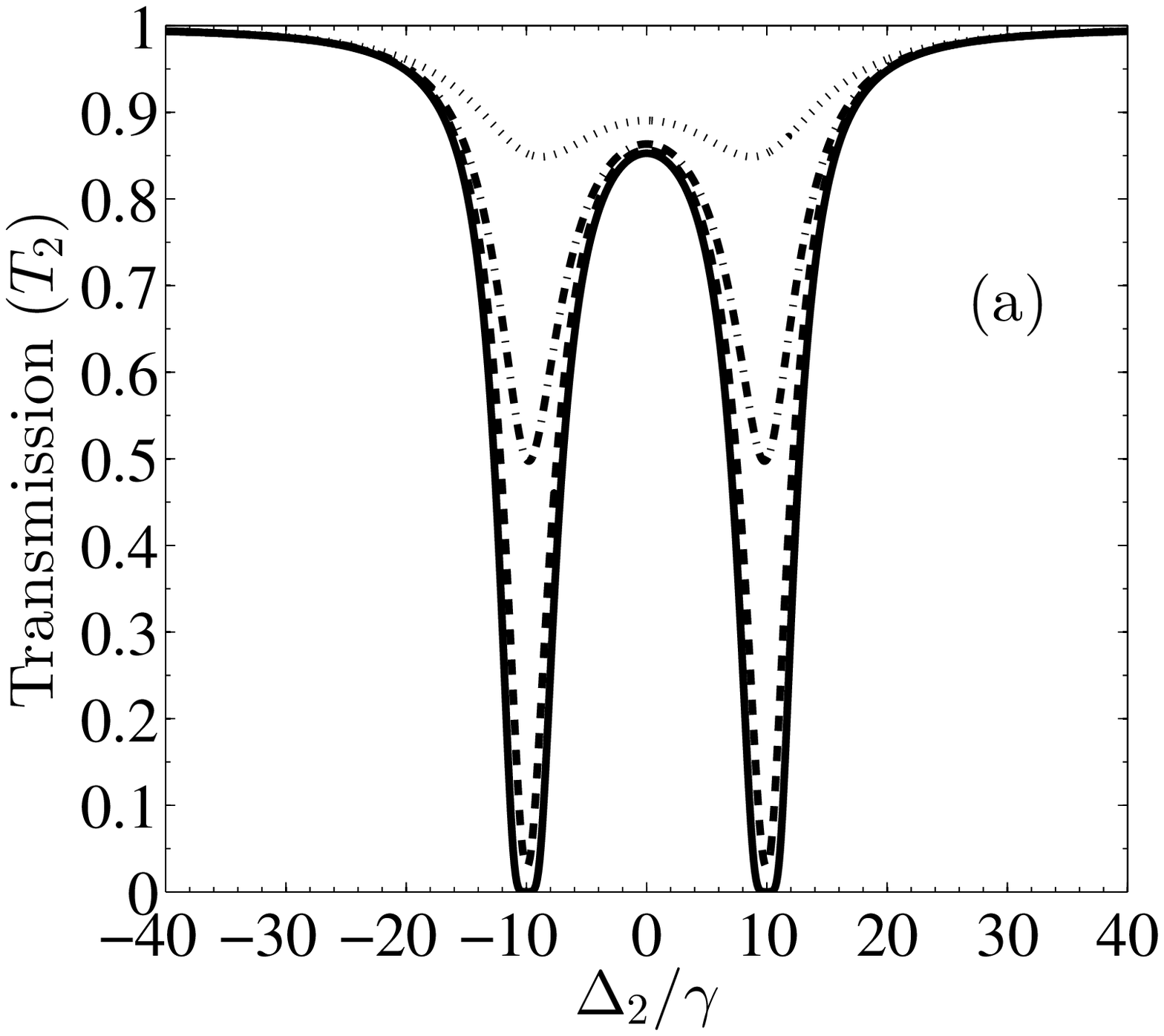}\\  \includegraphics[scale=0.4]{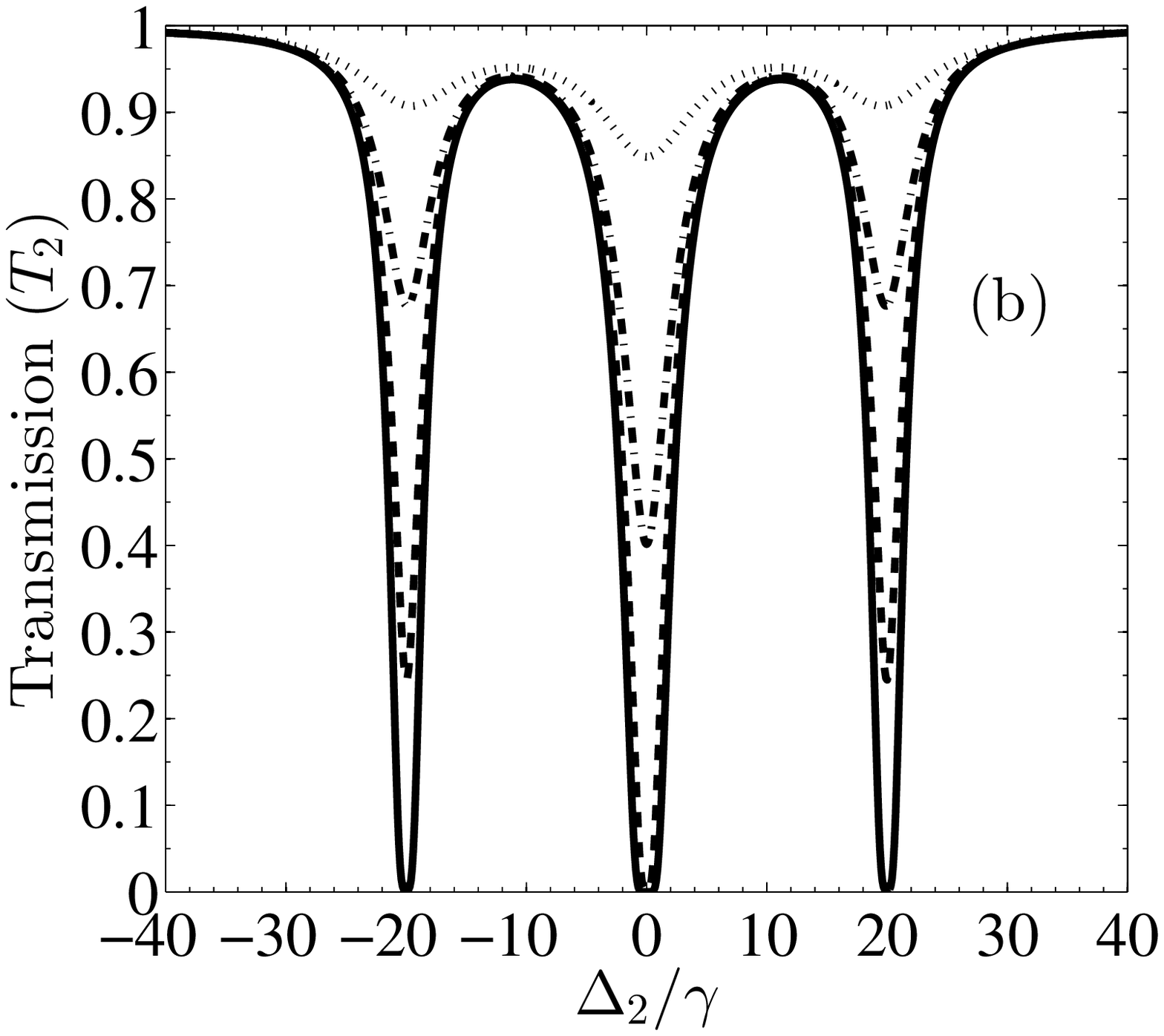}\\
\end{tabular}
\end{center}
\caption{Variation of transmission $\left(T_{2}\right)$ with detuning $\left(\Delta_{2}/\gamma\right)$ for (a) $G_{1}=10\gamma$, $G=0$ and (b) $G_{1}=10\gamma$, $G=10\gamma$. The common parameters of the above graphs are  $G_{2}=\gamma$ (solid line), $G_{2}=3\gamma$ (dashed line), $G_{2}=5\gamma$ (dot-dashed line), and $G_{2}=10\gamma$ (dotted line), $\Delta_{1}=0$, $\Delta=0$,  $\mathcal{N}=10^{10}~\mbox{cm}^{-3}$, $\eta_{12}=12$, $\eta_{23}=16$, $\eta_{34}=0.2$, $\gamma_{coll}=0$. }
\label{fig2}
\end{figure}
In order to describe the transmission of the field $\vec{E}_{2}$, the field equations [Eq. (\ref{eq3})] are solved in the steady-state limit, i.e., $\left(\partial G_{i}/\partial t\right)=\left(\partial \tilde{\rho}_{ij}/\partial t\right)=0$. We do not make any approximation on the strength of the fields, so we resorted to solve the set of simultaneous coupled equations numerically to all orders of these fields. To obtain the polarization, we solve Eq. (\ref{eqa1}) (in Appendix) in steady state and then integrate Eq. (\ref{eq3}) over the length of the medium. The absorption spectrum of the probe field is shown in Fig. \ref{fig2}. In the absence of the control field $\left(G=0\right)$, clearly, a transparency window appears around resonance [Fig. 2(a)], that pertains to EIT in three-level system in ladder configuration. For lower strength of the field $\vec{E}_{2}$ (viz., $G_{2}=\gamma$), the Aulter-Townes absorption doublet also arises at $\Delta_{2}=\pm G_{1}$, that can be attributed to single-photon absorption from dressed states $|\pm\rangle_{12}$ to $|3\rangle$, where $|\pm\rangle_{12}=\frac{1}{\sqrt{2}}\left(|1\rangle\pm|2\rangle\right)$ are the partial dressed states created by the resonant coupling field. The destructive interference between the transitions $|3\rangle\leftrightarrow|+\rangle_{12}$ and $|3\rangle\leftrightarrow|-\rangle_{12}$ gives rise to this transparency at resonance (Note that this transparency window does not correspond to zero absorption, due to the decay channel $|3\rangle\rightarrow |2\rangle$).

For \textit{larger} value of $G_2$, the transparency becomes prominent at resonance as the coherent absorption dominates the decay of the level $|3\rangle$. It is to be emphasized that the increase in transparency with the increase of the probe field is a consequence of the incoherent population distribution among various levels at the steady state. 

When the control field is introduced on $|3\rangle\leftrightarrow |4\rangle$ transition, (viz. $G=G_{1}$), the system becomes highly absorptive. While the new absorption peak appears at resonance (at $\Delta_{2}=\pm\left(G-G_{1}\right)$)  for smaller values of $G_{2}=\gamma$ [Fig. \ref{fig2}(b)], the control field also shifts the absorption doublet to $\Delta_{2}=\pm\left(G+G_{1}\right)$.  These absorption characteristics can be understood in terms of different partial dressed states \cite{mulchan2000}. The highest central absorption peak at $\Delta_{2}=\pm\left(G-G_{1}\right)$ i.e. at resonance arises due to the quantum interference between the transitions $|+\rangle_{34}\leftrightarrow|+\rangle_{12}$ and $|-\rangle_{34}\leftrightarrow|-\rangle_{12}$, where $|\pm\rangle_{34}=\frac{1}{\sqrt{2}}\left(|3\rangle\pm|4\rangle\right)$ are the dressed states produced by the control field $G$.    
On the other hand, the lower absorption peaks at $\Delta_{2}=\pm\left(G+G_{1}\right)$ are the consequences of the transitions  $|+\rangle_{34}\leftrightarrow|-\rangle_{12}$ and $|-\rangle_{34}\leftrightarrow|+\rangle_{12}$, respectively.   For larger value of $G_2$, the resonant absorption in the $|2\rangle\rightarrow |3\rangle$ transition becomes less again due to incoherent population distribution (leading to saturation).

\begin{figure}[!ht]
\begin{center}
\includegraphics[scale=0.4]{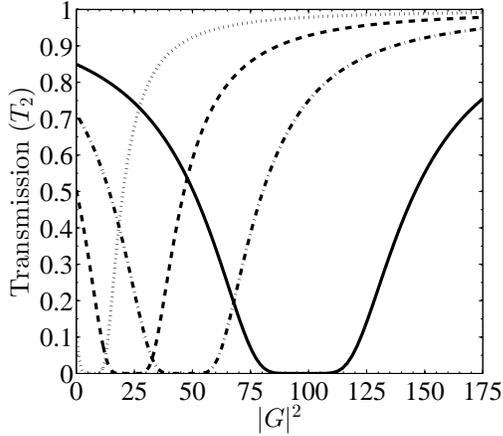}
\end{center}
\caption{Variation of transmission $\left(T_{2}\right)$ with intensity of control field  $\left(|G|^{2}\right)$ for $G_{1}=3\gamma$ (dotted line), $G_{1}=5\gamma$ (dashed line), $G_{1}=7\gamma$ (dot-dashed line), and $G_{1}=10\gamma$ (solid line). The other parameters are $G_{2}=0.01\gamma$, $\Delta_{2}=0$, and the other parameters are the same as in Fig. \ref{fig2}.}
\label{fig3}
\end{figure}

As discussed above, for $G=G_{1}$ and smaller values of $G_{2}$, the probe field $G_2$ gets fully absorbed at resonance. In support of this, we display in Fig. \ref{fig3} the transmission of resonant probe field as a function of the intensity of control field. It immediately becomes obvious that transmission vanishes at $G=G_{1}$. For $G<G_{1}$ the absorption dominates with increase of the intensity of the control field while  in the region $G>G_{1}$, the transmission increases with increase of control field. Thus, the condition $G=G_{1}$ is suitable to produce nonlinear absorptive switching because the probe absorption can be turned on and off by the control field. It should be borne in mind that such a large absorption at resonance for a weaker probe field occurs due to quantum interference enhanced nonlinear absorption with the inhibition of linear absorption. 

\begin{figure}[!ht]
\begin{center}
\begin{tabular}{cc}
\includegraphics[scale=0.4]{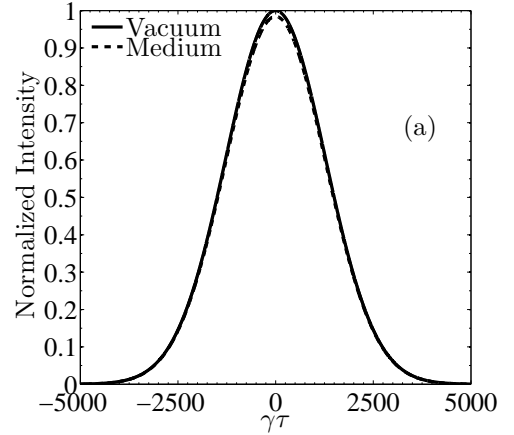}\\
\includegraphics[scale=0.4]{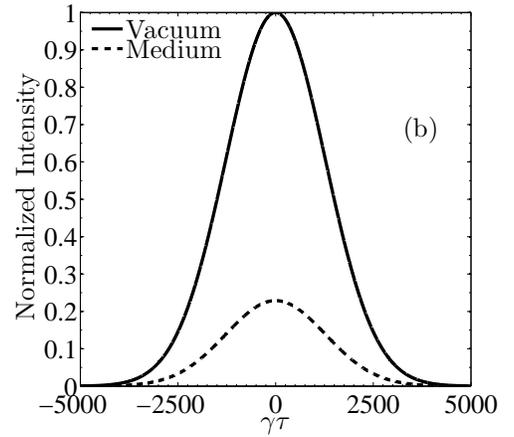}\\
\end{tabular}
\end{center}
\caption{Time dependence of the normalized Gaussian pulse after traveling through vacuum (solid line) and the atomic medium (dotted line) of length $L=1$ cm for (a) $G=0$ and (b) $G=10\gamma$. The parameters chosen are $G_{1}=10\gamma$, $G_{2}=0.1\gamma$, $\sigma=2\pi\times 5$ kHz, $\tau=t-L/c$ and other parameters are the same as in Fig. \ref{fig2}.}
\label{fig4}
\end{figure}
We now demonstrate the above analysis in terms of the absorptive photon switching with reference to a Gaussian probe pulse with a normalized envelope given by \cite{kumar2015}
\begin{align}\label{eq4}
\varepsilon\left(\omega\right)&=\varepsilon_{0}\frac{1}{\sigma\sqrt{\pi}}\mbox{exp}\left[-\omega^{2}/\sigma^{2}\right]\;,\nonumber\\
\varepsilon\left(t\right)&=\varepsilon_{0}\frac{1}{\sqrt{2\pi}}\mbox{exp}\left(-\sigma_{t}^{2}t^{2}/4\right)\;.
\end{align}
where $\sigma\left(\sigma_{t}=2/\sigma\right)$ denotes the width of the pulse in the frequency (time) domain and $\varepsilon_{0}$ is the pulse amplitude. For numerical calculation, we choose $\sigma=2\pi\times 5$ kHz ($\sigma_{t}=64~ \mu$s). In Fig. \ref{fig4}, we display the effect of control field on the absorption of a Gaussian shaped probe pulse Eq. (\ref{eq4}). In the absence of the control field, the output pulse remains equally intense as that of input pulse, thanks to EIT created by the coupling field of Rabi frequency $G_{1}=10\gamma$ as shown in Fig. \ref{fig4}(a). However, significant decrease in the output intensity occurs in the presence of the control field of Rabi frequency $G=G_{1}$ as depicted in Fig. \ref{fig4}(b).   Thus, the probe pulse can be turned on and off by the control field which demonstrates realization of the nonlinear absorptive photon switching \cite{harris1998,yan2001}.

\subsection{Control of Optical Bistability}
\begin{figure}[!ht]
\begin{center}
\includegraphics[scale=0.22]{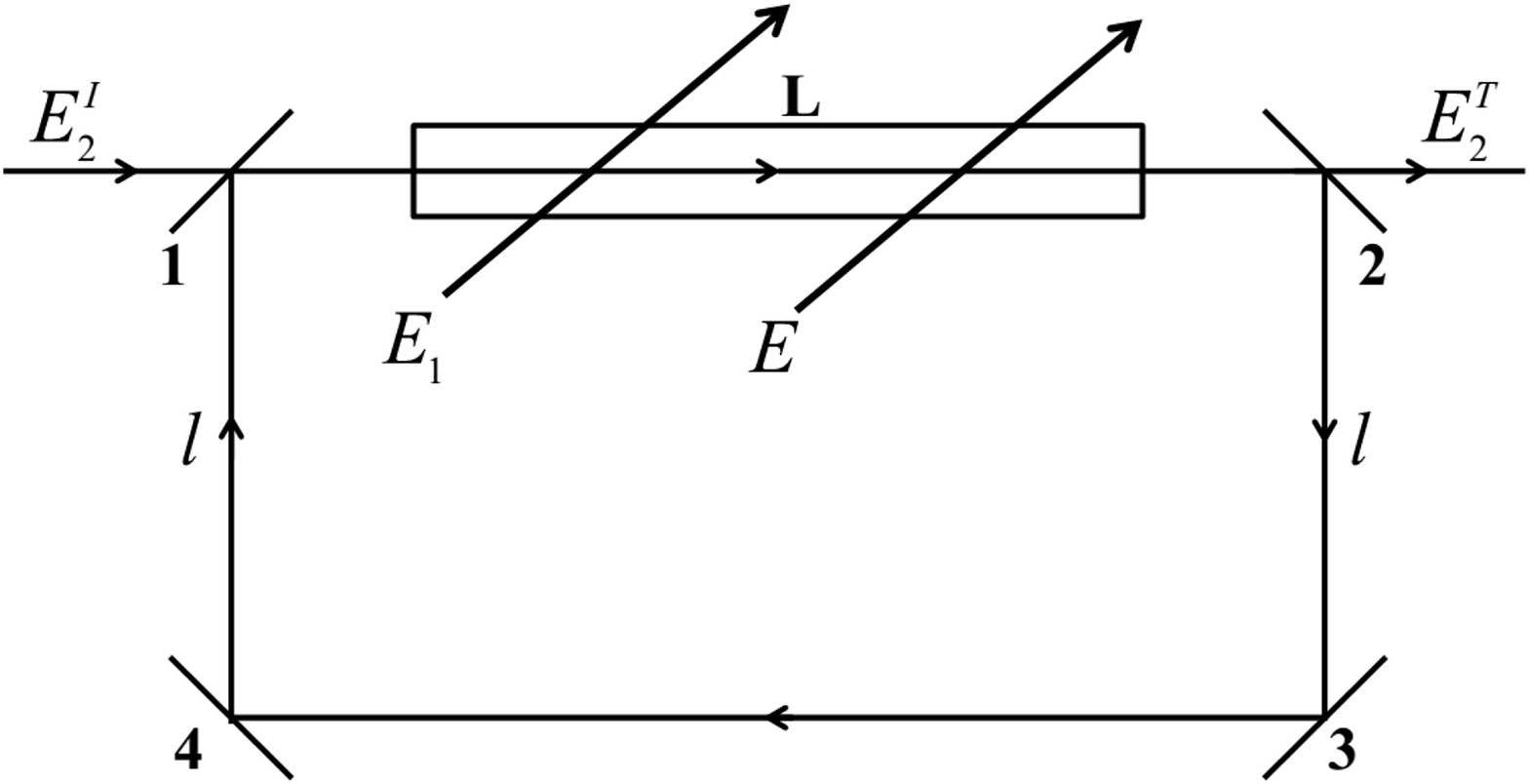}
\end{center}
\caption{Unidirectional ring cavity with the mirrors 3 and 4 as perfect reflectors, while the mirrors 1 and 2 obey $R+T=1$. The atomic medium of length $L$ is inserted in the cavity. $E_{2}^{I}$ and $E_{2}^{T}$ are the incident and transmitted fields and $\vec{E}$ and $\vec{E}_{1}$ are the control and coupling fields, respectively.}
\label{fig5}
\end{figure}
Now, we consider an optical feedback scenario, in which, an ensemble of $\mathcal{N}$ atoms in four-level ladder configuration [Fig. \ref{fig1}] is placed in a unidirectional cavity of total length $L_{T}$ as depicted in Fig. \ref{fig5}. Here, mirrors 3 and 4 are perfect reflectors, while $R$ and $T (=1-R)$ are the reflection and transmission coefficients of the semi-silvered mirrors 1 and 2. The total electric field experienced by the atoms placed in the cavity can be written as
\begin{align}\label{eq5}
\vec{E}=\vec{\varepsilon}_{1}e^{-i\omega_{G_{1}}t}+\vec{\varepsilon}_{2}e^{-i\omega_{G_{2}}t}+\vec{\varepsilon}_{c}e^{-i\omega_{G}t}+ c.c.\;,
\end{align}
The coherent field $E_{2}^{I}$ enters the cavity through the semi-silvered mirror 1 and induces polarization [Eq. (\ref{eq2})] in the medium. This field comes out of the cavity and gets partially transmitted from the mirror 2 as $E_{2}^{T}$. The fields $\vec{E}_{1}$ and $\vec{E}$ do not circulate in the cavity; but these fields control the induced polarization. For a perfectly tuned cavity, the boundary conditions  between the incident field $E_{2}^{I}$ and the transmitted field $E_{2}^{T}$ are given by \cite{bonifacio1976,bonifacio1978}
\begin{center}
\begin{align}\label{eq6}
E_{2}^{T}(t)&=\sqrt{T}E_{2}\left(L,t\right)\;,\nonumber\\
E_{2}(0,t)&=\sqrt{T}E_{2}^{I}(t)+R~\mbox{exp}\left(-i\delta_{0}\right)E_{2}\left(L,t-\Delta t\right)\;,
\end{align}
\end{center}
where, $E_{2}(0,t)$ is the field at the start of the sample, $E_{2}(L,t)$ is the field after traversing the sample of length $L$, $\Delta t=\left(2l+L\right)/c$ is the time taken to travel from mirror 2 to mirror 1, $\delta_{0}=\left(\omega_{c}-\omega_{0}\right)L_{T}/c$ is the detuning with the cavity with frequency $\omega_{c}$, and $L_{T}=2\left(l+L\right)$ is the total length of the cavity. In the steady state, the above boundary conditions reduce to
\begin{align}\label{eq7}
E_{2}^{T}&=\sqrt{T}E_{2}\left(L\right)\;,\nonumber\\
E_{2}(0)&=\sqrt{T}E_{2}^{I}+R~\mbox{exp}\left(-i\delta_{0}\right)E_{2}\left(L\right)\;.
\end{align}
We define a cooperation parameter $C=\alpha L/2T$ \cite{abraham1982,lugiato1984}, where, $\alpha=\frac{4\pi\mathcal{N}\omega_{G_{2}}|\vec{d}_{32}|^{2}}{\hbar \gamma_{32}c}$ is the absorption coefficient  on the transition $|2\rangle\leftrightarrow|3\rangle$.
\begin{figure}[!ht]
\begin{center}
\begin{tabular}{cc}
\includegraphics[scale=0.4]{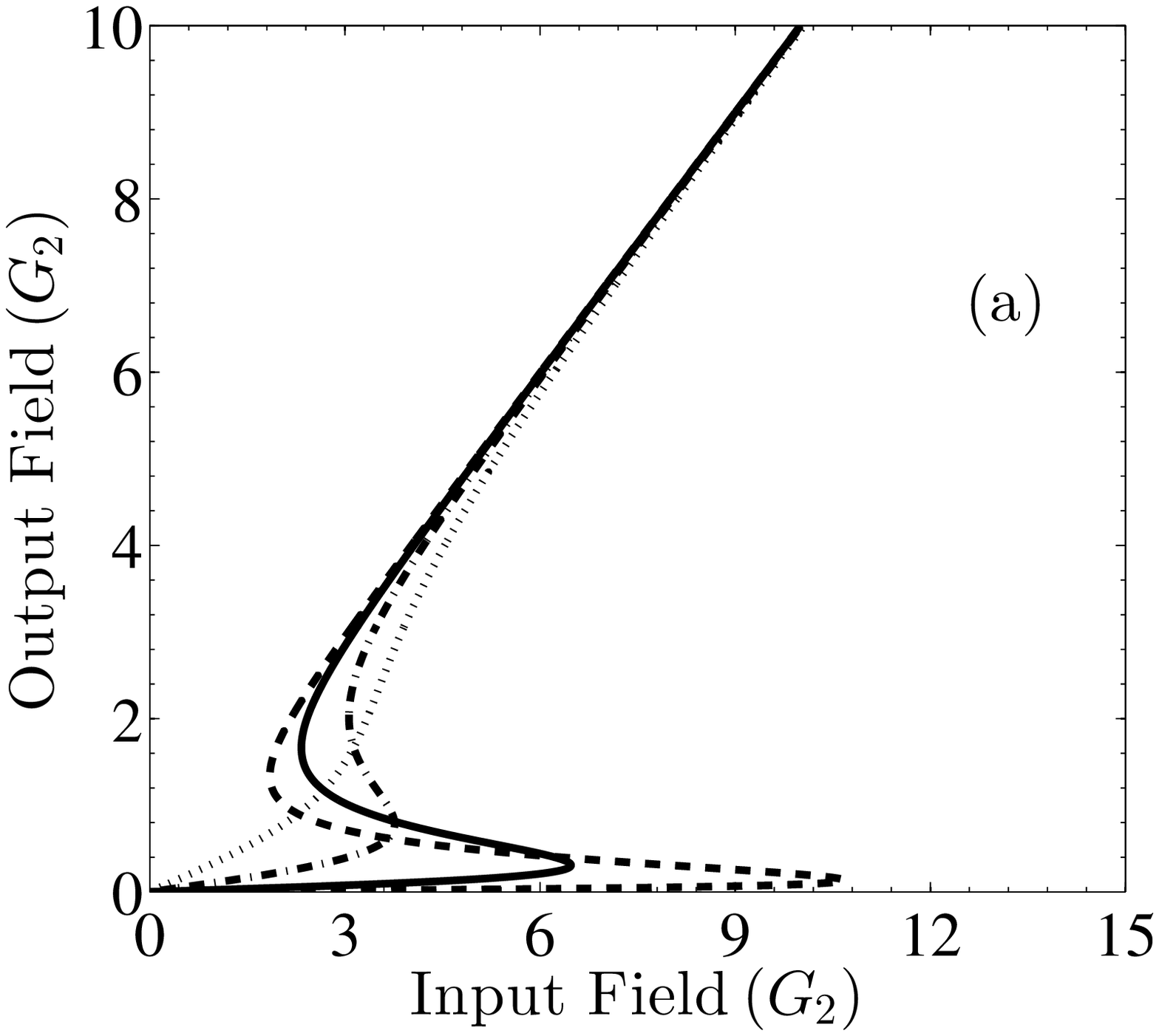}\\
\includegraphics[scale=0.4]{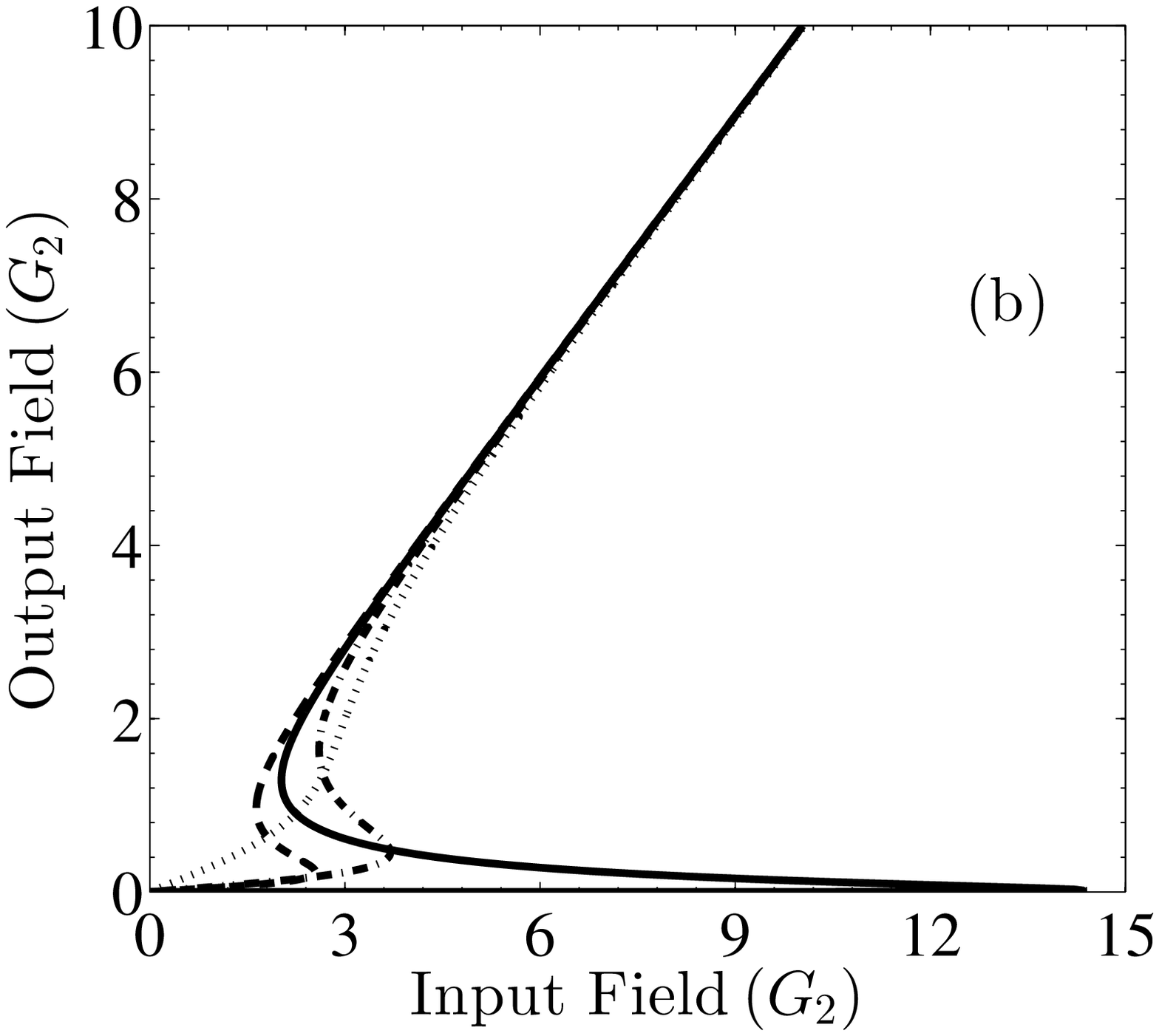}\\
\end{tabular}
\end{center}
\caption{Variation of output fields with input fields for (a) $G=0$ and (b) $G=5\gamma$. The common parameters for above graphs are $G_{1}=3\gamma$ (dashed line), $G_{1}=5\gamma$ (solid line), $G_{1}=10\gamma$ (dot-dashed line), and $G_{1}=20\gamma$ (dotted line). The parameters are chosen as $\Delta_{2}=0$, $C=400$ and the other parameters are the same as in Fig. \ref{fig2}.}
\label{fig6}
\end{figure}

Fig. \ref{fig6}(a) shows the dependence of the optical bistability on the input laser intensity $G_{2}$ in the absence of the control field ($G=0$) for different values of $G_{1}$ without mean-field approximation \cite{harshawardhan1996}. In absence of the control field, the system acts as a three-level ladder system, in which the increase of $G_{1}$ leads to transparency of $G_{2}$, in analogy to EIT. As seen in Fig. \ref{fig6}(a), the threshold of the optical bistability decreases with increase of the input field $G_{1}$. For much larger values of $G_{1}=20\gamma$, the system actually shows the features of optical transistor. 

When the control field is \textit{switched on}, absorption dominates at resonance for $G=G_{1}$. Despite this, the region for $G\lessgtr G_{1}$ corresponds to transparency. As depicted in Fig. \ref{fig6}(b), $G=G_{1}$ leads to significant increase of the threshold. On the other hand, for $G\neq G_{1}$, the threshold for bistability remains quite small. Thus, by working in the region $G \lessgtr G_{1}$, the  nonlinearity of the atomic medium can be greatly enhanced and hence it is easier to achieve the saturation for the cavity field, than in the case when the control field is not applied. In fact, in absence of the control field, one needs to apply a very large coupling field to achieve smaller threshold [see Fig. \ref{fig6}(a)].

\section{Control of RSA and SA: Effect of decay rates}
\begin{figure}[!ht]
\begin{center}
\includegraphics[scale=0.2]{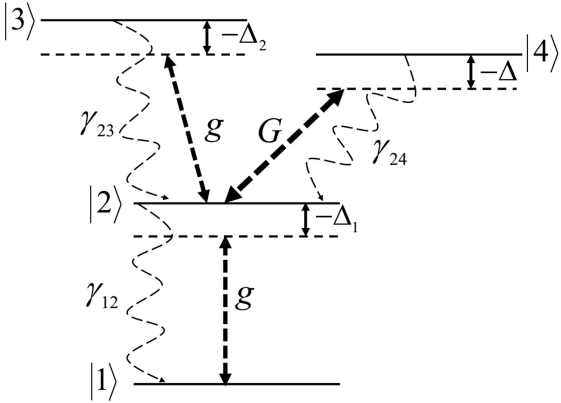}
\end{center}
\caption{Level scheme for the SA and RSA effects. The two dipole allowed transitions $|1\rangle\leftrightarrow|2\rangle$ and $|2\rangle\leftrightarrow|3\rangle$ are coupled to the fields having same polarization and Rabi frequency 2$g$. The control field drives the transition $|2\rangle\leftrightarrow|4\rangle$.}
\label{fig7}
\end{figure}
Next, to model the SA and RSA effects, we couple the first excited state of three-level ladder system with a metastable state by a control field, so as to form a Y-type system as shown in Fig. \ref{fig7}. For the purpose of SA and RSA effects,  we choose the fields $\vec{E}_{1}$ and $\vec{E}_{2}$  to be of the same polarization and of equal Rabi frequency 2$g$. From now onwards, we call this field as a \textit{probe field} which is interacting with transitions $|1\rangle\leftrightarrow|2\rangle$ and $|2\rangle\leftrightarrow|3\rangle$. Here, the control field of Rabi frequency 2$G$ is coupled to transition $|2\rangle\leftrightarrow|4\rangle$. The relevant density matrix equations for Y-type system which describe the dynamical evolution of the system by using Markovian master equation are given in Appendix B [see Eq. (\ref{eqb1})].

Thus, the polarization induced in the medium can  be written as
\begin{align}\label{eq8}
P\left(\omega\right)=\mathcal{N}\vec{d}\left(\tilde{\rho}_{21}+\tilde{\rho}_{32}\right)\;.
\end{align}
where, we have assumed that $\vec{d}=\vec{d}_{12}=\vec{d}_{23}$. Further, the Maxwell's equations in slowly varying envelope approximation can be written as
\begin{align}\label{eq9}
\left(\frac{\partial}{\partial z}+\frac{1}{c}\frac{\partial}{\partial t}\right)g &=i\left(\eta_{12}\tilde{\rho}_{21}+\eta_{23}\tilde{\rho}_{32}\right)\;,\nonumber\\
\left(\frac{\partial}{\partial z}+\frac{1}{c}\frac{\partial}{\partial t}\right)G &=i\eta_{24}\tilde{\rho}_{43}\;.
\end{align}
where $\eta_{ij}$ represents the coupling constant for $|i\rangle\leftrightarrow|j\rangle$ transition. To obtain transmission at the output of the medium, we integrate Eq. (\ref{eq9}) over the length of the medium in steady state. 
\begin{figure}[!ht]
\begin{center}
\begin{tabular}{cc}
\includegraphics[scale=0.4]{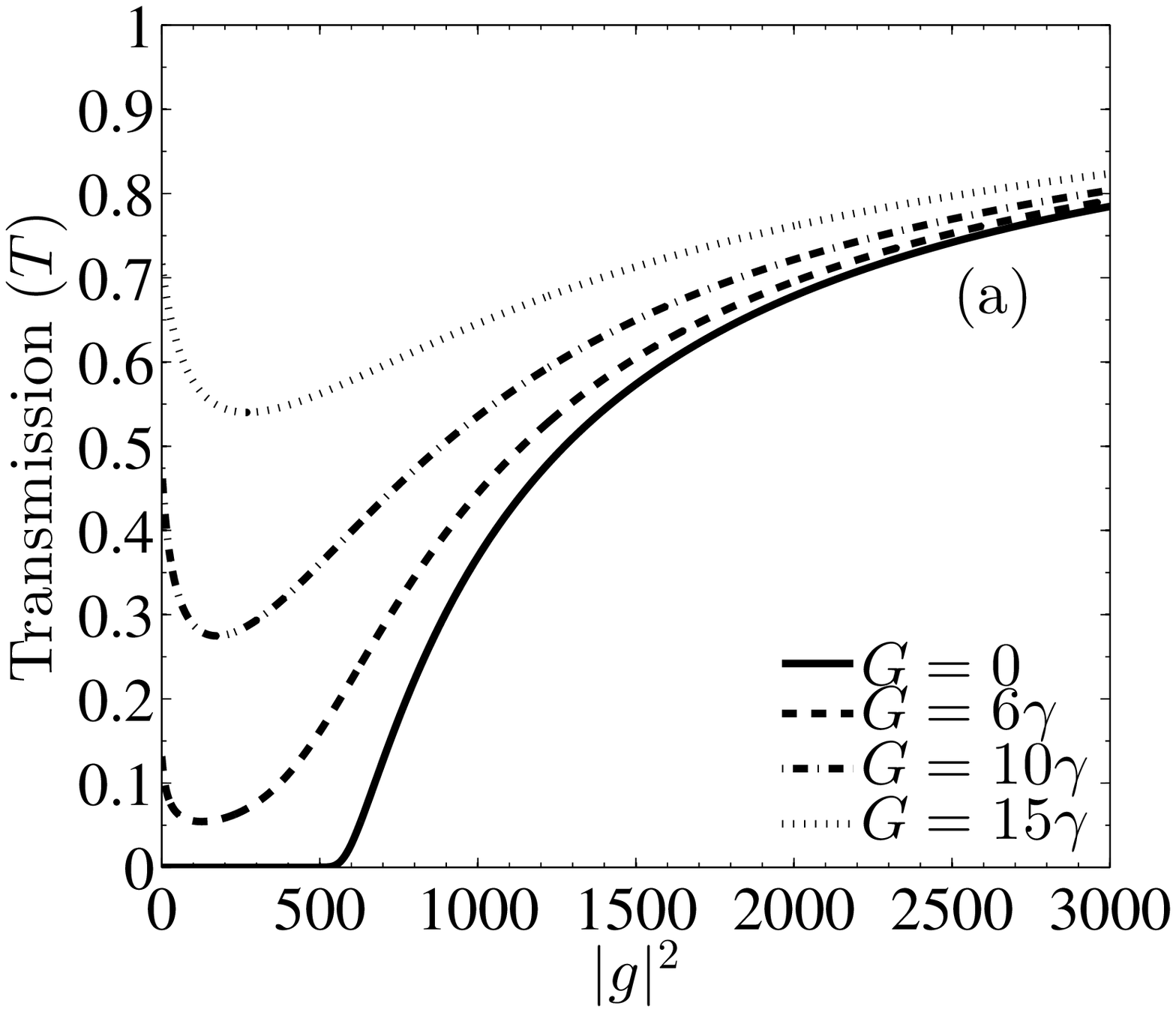}\\
\includegraphics[scale=0.4]{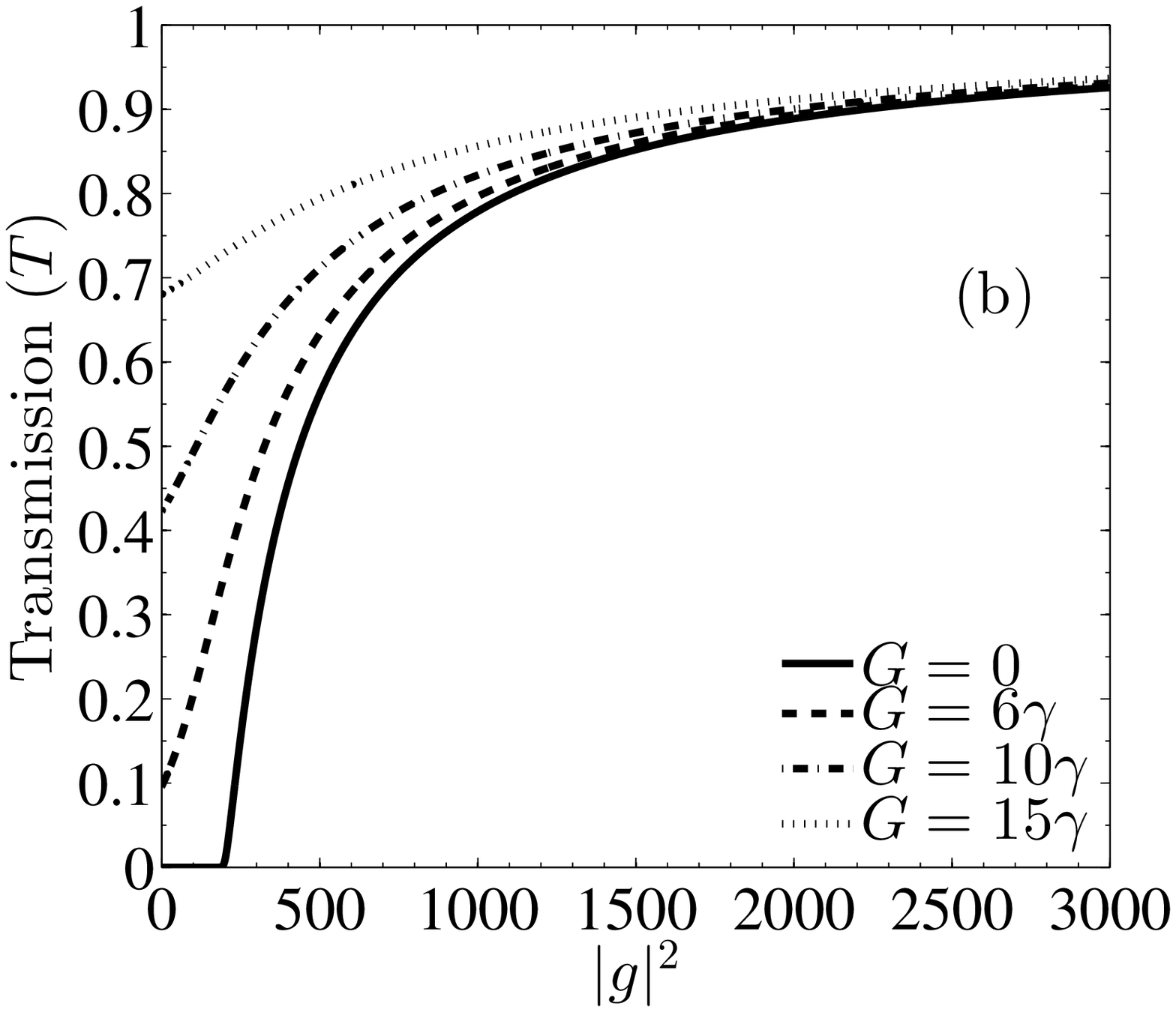}\\
\end{tabular}
\end{center}
\caption{Variation of transmission $\left(T\right)$ with the intensity of the probe field for (a)  $\gamma_{12}/2\pi=$ 5 MHz, $\gamma_{23}/2\pi=$ 11 MHz, $\gamma_{24}/2\pi=$ 0.97 MHz, $\eta_{12}=88$, $\eta_{23}=1.5$, and $\eta_{24}=8.8$, (b) $\gamma_{12}/2\pi=$ 6 MHz, $\gamma_{23}/2\pi=$ 0.97 MHz,  $\gamma_{24}/2\pi=$ 1.1 MHz, $\eta_{12}=87$, $\eta_{23}=14$, and $\eta_{24}=10$. Here the fields are resonant, i.e., $\Delta_{1}=0$, $\Delta_{2}=0$, $\Delta=0$ and we have chosen $\gamma=$1 MHz.}
\label{fig8}
\end{figure}

We first consider that the level $|3\rangle$ has a decay rate larger than the level $|2\rangle$. We show in Fig. \ref{fig8}(a) how the net transmission of the input field varies as a function of the input intensity. We find that as the input intensity increases upto a certain threshold, the system remains absorptive, in absence of the control field. This is due to the fact that the level $|3\rangle$ decays much faster. At a certain input threshold, when the input intensity dominates over this decay rate, the system undergoes saturation and exhibits more transmission with larger input intensity, as in the case of SA.  Next when a control field is introduced, we find that the system becomes transparent for weak probe field, due to the EIT-like effect. This phenomenon can be understood as a manifestation of the partial dressed states that comprise of the levels $|2\rangle$ and $|4\rangle$. As the input intensity is increased until the input threshold, the system exhibits more absorption, leading to the RSA effect. This is because the control field assists in further absorption by Rabi cycling the population in the ($|2\rangle, |4\rangle$) manifold. \textit{Clearly, just by applying a control field, one can achieve RSA effect in a system that exhibits SA otherwise}. We illustrate the RSA effect by considering the example of the relevant energy levels of $^{85}$Rb as: $|1\rangle$= 5$S_{1/2}$, $|2\rangle$= 5$P_{1/2}$, $|3\rangle$= 7$S_{1/2}$ and $|4\rangle$= 5$D_{3/2}$. The respective transition wavelengths and decay rates are \cite{wielandy1998,rong2009} $\lambda_{12}= $795 nm and $\gamma_{12}= 2\pi\times$5 MHz (for $|1\rangle\leftrightarrow|2\rangle$ transition), $\lambda_{23}=$ 762 nm and $\gamma_{23}=2\pi\times$11 MHz (for $|2\rangle\leftrightarrow|3\rangle$ transition), $\lambda_{24}=$ 741 nm and $\gamma_{24}=2\pi\times$0.67 MHz (for $|2\rangle\leftrightarrow|4\rangle$ transition). For RSA effect, $\hat{\pi}$-polarized field is interacting with $|1\rangle\leftrightarrow|2\rangle$ and $|2\rangle\leftrightarrow|3\rangle$ transitions whereas the control field is $\hat{\sigma}_{+}$-polarized on $|2\rangle\leftrightarrow|4\rangle$ transition.

Next we consider a situation in which the level $|3\rangle$ is slowly decaying compared to the level $|2\rangle$. In this case, as the population resides in the level $|3\rangle$ for much longer time, the system achieves saturation quickly and exhibits transparency (and SA). When a control field is introduced, this advances the saturation. This is because, the level $|4\rangle$ gets populated through coherent coupling to the level $|2\rangle$ and thus assists in transmission.  In Fig. \ref{fig8}(b), we display this effect of the control field on SA. For the possible realization of SA, we consider the relevant energy levels of $^{87}$Rb. The designated states can be chosen as follows: $|1\rangle$= 5$S_{1/2}$, $|2\rangle$= 5$P_{3/2}$, $|3\rangle$= 5$D_{5/2}$ and $|4\rangle$= 8$S_{1/2}$. The respective transition wavelengths and decay rates are \cite{yong1995,olson2009,safronova2011} $\lambda_{12}=$ 780 nm and $\gamma_{12}=2\pi\times$6 MHz (for $|1\rangle\leftrightarrow|2\rangle$ transition), $\lambda_{23}=$ 776 nm and $\gamma_{23}=2\pi\times$0.97 MHz (for $|2\rangle\leftrightarrow|3\rangle$ transition), $\lambda_{24}=$ 616 nm and $\gamma_{24}=2\pi\times$1.1 MHz (for $|2\rangle\leftrightarrow|4\rangle$ transition). For SA effect, $\hat{\sigma}_{+}$-polarized field is interacting with $|1\rangle\leftrightarrow|2\rangle$ and $|2\rangle\leftrightarrow|3\rangle$ transitions while the polarization of the control field is $\hat{\sigma}_{-}$ on $|2\rangle\leftrightarrow|4\rangle$ transition.
\vspace{-0.54cm}
\section{Conclusions}
In conclusions, we have demonstrated the possibility of coherent control of nonlinear absorption of a probe field in four-level atomic systems. We have shown that in a four-level ladder system, the medium which otherwise is transmissive for a weaker probe field, becomes absorptive as the control field is introduced. Such a large nonlinear absorption leads to the absorptive optical switching. Further, as the strength of the probe field is increased, saturation effect gives rise to large transparency.  We further have shown that the threshold for optical bistability can be manipulated both by coupling field and the control field. Further, in a four level Y-configuration, we have discussed how the control field can work as a knob to switch from RSA into SA.  

\appendix
\section{Density matrix equations for four level ladder system}
To describe the dynamics of the four-level ladder system, we use Markovian master equation and obtain the following density matrix equations:
\begin{align}\label{eqa1}
\dot{\tilde{\rho}}_{22} &= \gamma_{23}\tilde{\rho}_{33}-\gamma_{12}\tilde{\rho}_{22}+i(G_{1}\tilde{\rho}_{12}-G_{1}^{\ast}\tilde{\rho}_{21})-i(G_{2}\tilde{\rho}_{23}-G_{2}^{\ast}\tilde{\rho}_{32})\nonumber\\
\dot{\tilde{\rho}}_{33} &= \gamma_{34}\tilde{\rho}_{44}-\gamma_{23}\tilde{\rho}_{33}+i(G_{2}\tilde{\rho}_{23}-G_{2}^{\ast}\tilde{\rho}_{32})-i(G\tilde{\rho}_{34}-G^{\ast}\tilde{\rho}_{43}) \nonumber\\
\dot{\tilde{\rho}}_{44} &= -\gamma_{34}\tilde{\rho}_{44}+i(G\tilde{\rho}_{34}-G^{\ast}\tilde{\rho}_{43}\nonumber)\\
\dot{\tilde{\rho}}_{21} &= \left(i\Delta_{1}-\Gamma_{21}\right)\tilde{\rho}_{21}+iG_{1}\left(1-2\tilde{\rho}_{22}-\tilde{\rho}_{33}-\tilde{\rho}_{44}\right)+iG_{2}^{\ast}\tilde{\rho}_{31}\nonumber\\
\dot{\tilde{\rho}}_{32} &= \left(i\Delta_{2}-\Gamma_{32}\right)\tilde{\rho}_{32}+iG_{2}\left(\tilde{\rho}_{22}-\tilde{\rho}_{33}\right)+iG^{\ast}\tilde{\rho}_{42}-iG_{1}^{\ast}\tilde{\rho}_{31}\nonumber\\
\dot{\tilde{\rho}}_{43} &= \left(i\Delta-\Gamma_{43}\right)\tilde{\rho}_{43}+iG\left(\tilde{\rho}_{33}-\tilde{\rho}_{44}\right)-iG_{2}^{\ast}\tilde{\rho}_{42}\nonumber\\
\dot{\tilde{\rho}}_{31} &= \left[i\left(\Delta_{1}+\Delta_{2}\right)-\Gamma_{31}\right]\tilde{\rho}_{31}+iG^{\ast}\tilde{\rho}_{41}+iG_{2}\tilde{\rho}_{21}-iG_{1}\tilde{\rho}_{32}\nonumber\\
\dot{\tilde{\rho}}_{42} &= \left[i\left(\Delta_{2}+\Delta\right)-\Gamma_{42}\right]\tilde{\rho}_{42}+iG\tilde{\rho}_{32}-iG_{2}\tilde{\rho}_{43}-iG_{1}^{\ast}\tilde{\rho}_{41}\nonumber\\
\dot{\tilde{\rho}}_{41} &= \left[i\left(\Delta_{1}+\Delta_{2}+\Delta\right)-\Gamma_{41}\right]\tilde{\rho}_{41}+iG\tilde{\rho}_{31}-iG_{1}\tilde{\rho}_{42}\;,
\end{align}
together with the condition $\sum\limits_{i=1}^4\tilde{\rho}_{ii}=1$.
Here, $\Delta_{1}=\omega_{G_{1}}-\omega_{21}$ $\left(\Delta_{2}=\omega_{G_{2}}-\omega_{32} \right)$ is the detuning of the field $\vec{E}_{1}$ $\left(\vec{E}_{2}\right)$ from the transition $|1\rangle\leftrightarrow|2\rangle$ $\left(|2\rangle\leftrightarrow|3\rangle\right)$ and $\Delta=\omega_{c}-\omega_{43}$ is the detuning of the control field from the transition $|3\rangle\leftrightarrow|4\rangle$. The   spontaneous emission rate from the level $|j\rangle$ to $|i\rangle$ is represented by $\gamma_{ij}$ and $\Gamma_{ij}=\frac{1}{2}\sum\limits_{k}\left(\gamma_{ki}+\gamma_{kj}\right)+\gamma_{coll}$ describe the dephasing rate of coherence between the levels $|j\rangle$ and $|i\rangle$, $\gamma_{coll}$ being the collisional decay rate. The highly oscillating terms in Eq. (\ref{eqa1}) are neglected with rotating wave approximations, by choosing the transformations for the density matrix elements
as: $\tilde{\rho}_{21}=\rho_{21}~e^{i\omega_{G_{1}}t}$, $\tilde{\rho}_{32}=\rho_{32}~e^{i\omega_{G_{2}}t}$, $\tilde{\rho}_{43}=\rho_{43}~e^{i\omega_{c}t}$,$~\tilde{\rho}_{31}=\rho_{31}~e^{i\left(\omega_{G_{1}}+\omega_{G_{2}}\right)t}$, $\tilde{\rho}_{42}=\rho_{42}~e^{i\left(\omega_{c}+\omega_{G_{2}}\right)t}$,$~\tilde{\rho}_{41}=\rho_{41}~e^{i\left(\omega_{c}+\omega_{G_{1}}+\omega_{G_{2}}\right)t}$, $\tilde{\rho}_{ii}=\rho_{ii}$.
\section{Density matrix equations for Y-type system}
The equations of evolution of density matrix by using Markovian master equation for Y-type system are as follows:
\begin{align}\label{eqb1}
\dot{\tilde{\rho}}_{11}&=\gamma_{12}\tilde{\rho}_{22}-i\left(g_{1}\tilde{\rho}_{12}-g_{1}^{\ast}\tilde{\rho}_{21}\right)\nonumber\\
\dot{\tilde{\rho}}_{22}&=-\gamma_{12}\tilde{\rho}_{22}+\gamma_{23}\tilde{\rho}_{33}+\gamma_{24}\tilde{\rho}_{44}
+i\left(g_{1}\tilde{\rho}_{12}-g_{1}^{\ast}\tilde{\rho}_{21}\right)\nonumber\\
&-i\left(g_{2}\tilde{\rho}_{23}-g_{2}^{\ast}\tilde{\rho}_{32}\right)-i\left(G\tilde{\rho}_{24}-G^{\ast}\tilde{\rho}_{42}\right)\nonumber\\
\dot{\tilde{\rho}}_{33} &=-\gamma_{23}\tilde{\rho}_{33}+i\left(g_{2}\tilde{\rho}_{23}-g_{2}^{\ast}\tilde{\rho}_{32}\right)\nonumber\\
\dot{\tilde{\rho}}_{44} &=-\gamma_{24}\tilde{\rho}_{44}-i\left(G\tilde{\rho}_{24}-G^{\ast}\tilde{\rho}_{42}\right)\nonumber\\
\dot{\tilde{\rho}}_{21} &=\left(i\Delta_{1}-\Gamma_{21}\right)\tilde{\rho}_{21}+ig_{1}\left(1-2\tilde{\rho}_{22}-\tilde{\rho}_{33}-\tilde{\rho}_{44}\right)\nonumber\\
&+ig_{2}^{\ast}\tilde{\rho}_{31}+iG^{\ast}\tilde{\rho}_{41}\nonumber\\
\dot{\tilde{\rho}}_{32} &=\left(i\Delta_{2}-\Gamma_{32}\right)\tilde{\rho}_{32}+ig_{2}\left(\tilde{\rho}_{22}-\tilde{\rho}_{33}\right)\nonumber\\
&-ig_{1}^{\ast}\tilde{\rho}_{31}+iG\tilde{\rho}_{34}\nonumber\\
\dot{\tilde{\rho}}_{42} &=\left(i\Delta-\Gamma_{42}\right)\tilde{\rho}_{42}+iG\left(\tilde{\rho}_{22}-\tilde{\rho}_{44}\right)\nonumber\\
&-ig_{2}^{\ast}\tilde{\rho}_{43}-ig_{1}^{\ast}\tilde{\rho}_{41}\nonumber\\
\dot{\tilde{\rho}}_{31} &=\left[i\left(\Delta_{1}+\Delta_{2}\right)-\Gamma_{31}\right]\tilde{\rho}_{31}-ig_{1}\tilde{\rho}_{32}+ig_{2}\tilde{\rho}_{21}\nonumber\\
\dot{\tilde{\rho}}_{43} &=\left[i\left(\Delta-\Delta_{2}\right)-\Gamma_{43}\right]\tilde{\rho}_{43}+iG\tilde{\rho}_{23}-ig_{2}^{\ast}\tilde{\rho}_{42}\nonumber\\
\dot{\tilde{\rho}}_{41}&=\left[i\left(\Delta_{1}+\Delta\right)-\Gamma_{41}\right]\tilde{\rho}_{41}+iG\tilde{\rho}_{21}-ig_{1}\tilde{\rho}_{42}
\end{align}
Here, we have used $\sum\limits_{i=1}^{4}\tilde{\rho}_{ii}=1$. The detunings of the fields from respective transitions are given by $\Delta_{1}=\omega_{G_{1}}-\omega_{12}$, $\Delta_{2}=\omega_{G_{2}}-\omega_{23}$ and $\Delta=\omega_{c}-\omega_{24}$. Further, the transformations used to remove the the rapidly oscillating terms in Eq. (\ref{eqb1}) are as: $\tilde{\rho}_{21}=\rho_{21}~e^{i\omega_{G_{1}}t}$, $\tilde{\rho}_{32}=\rho_{32}~e^{i\omega_{G_{2}}t}$, $\tilde{\rho}_{42}=\rho_{42}~e^{i\omega_{c}t}$, $\tilde{\rho}_{31}=\rho_{31}~e^{i\left(\omega_{G_{1}}+\omega_{G_{2}}\right)t}$, $\tilde{\rho}_{41}=\rho_{41}~e^{i\left(\omega_{G_{1}}+\omega_{c}\right)t}$, $\tilde{\rho}_{41}=\rho_{41}~e^{i\left(\omega_{c}-\omega_{G_{2}}\right)t}$, and $\tilde{\rho}_{ii}=\rho_{ii}$.
\bibliographystyle{elsarticl-num}

\end{document}